\documentclass{wbk-p10x7}

\begin{document}

\title{A Next-to-Leading Order Calculation of Hadronic Three Jet Production\footnote{}
}

\def\preprintno{BNL-HET-00/34}
\def\paperid{W. Kilgore (ICHEP 2000)}
\author{William B. Kilgore}

\address{Physics Department, Building 510A, Brookhaven National
Laboratory, Upton, NY 11973, USA\\E-mail: kilgore@bnl.gov}

\author{Walter T. Giele}

\address{Theoretical Physics Department, Fermi National Accelerator
Laboratory, Batavia, IL 60510, USA\\E-mail: giele@fnal.gov}

\twocolumn[\maketitle\abstract{
We present results of a next-to-leading order calculation of three jet 
production at hadron colliders.  This calculation will have many applications. 
In addition to computing such three-jet observables  as spectra, mass
distributions, this calculation permits the first next-to-leading
order studies at hadron colliders of jet and event shape variables.}]

\footnotetext{${}^*$Talk presented by W.B.K. at the XXXth International
Conference on High Energy Physics, Osaka, Japan, July 27 -- August 2, 2000.}

\section{Introduction}

One of the difficulties in interpreting experimental results is
in assessing the uncertainty to be associated with the theoretical
calculation.  This is particularly true in QCD where the coupling is
quite strong and one expects higher order corrections to be significant.

Typically, one characterizes theoretical uncertainty by the dependence
on the renormalization scale $\mu$. Since we don't actually know how
to choose $\mu$ or even a range of $\mu$, the uncertainty associated
with scale dependence is somewhat arbitrary.  One motivation for
performing next-to-leading order (NLO) calculations is to reduce the
scale dependence associated with the calculation.

However, this is not the only benefit of an NLO calculation.  There
are times when the LO calculation is a bad estimator of the
physical process.  It may be that leading order kinematics
artificially forbids the most important physical process.  It could
also be that the NLO corrections are simply large.  
Even if the overall NLO correction is relatively small, there may be
regions of phase space, where NLO corrections are large.  It is
only in those regions of phase space where the NLO corrections are
well behaved (as determined by the ratio of the NLO to LO terms) that
one has confidence in the reliability of the calculation and can begin
to believe the uncertainty estimated from scale dependence and it is
only when one has a reliable estimate of the theoretical uncertainty
that comparisons to experiment are meaningful.

\section{Methods}
The NLO three jet calculation consists of two parts: two to three
parton processes at one-loop ($gg\to ggg$\cite{BDK1},
$\overline{q}q\to ggg$\cite{BDK2},
$\overline{q}q\to\overline{Q}Qg$\cite{KST}, and processes related to 
these by crossing symmetry) and two to four parton processes ($gg\to
gggg$, $\overline{q}q\to gggg$, $\overline{q}q\to\overline{Q}Qgg$, and
$\overline{q}q\to\overline{Q}Q\overline{Q^\prime}Q^\prime$, and the
crossed processes) computed at tree-level.  Both of these contributions
are infrared singular; only the sum of the two is infrared finite and
meaningful.  The Kinoshita-Lee-Nauenberg theorem\cite{kln} guarantees
that the infrared singularities of the one-loop processes cancel those
of the real emission processes for sufficiently inclusive observables.

In order to implement the kinematic cuts necessary to compare a
calculation to experimental data one must compute the cross section
numerically.  Thus, we must find a numerically safe way of canceling
the singularities.  The method we use is the ``subtraction improved''
phase space slicing method\cite{GG,GGK,KG}.

\section{Applications}
The next-to-leading order calculation of three jet production will
have a wide array of phenomenological applications.
\subsection{Measurement of $\alpha_s$}
It should be possible to extract a purely hadronic measurement of
$\alpha_s$.  One possibility for such a measurement would be a
comparison of the three jet to two jet event rate\cite{Ratio}.  Since
both processes are sensitive to all possible initial states at
tree-level, a next-to-leading order comparison should be relatively
free of bias from the parton distributions.  Because the measurement
will be simultaneously performed over a wide range of energy scales,
the running of $\alpha_s$ can be used to constrain the fits and
enhance the precision of the combined measurement\cite{GGY}.

It has also been suggested that $\alpha_s$ can be determined from
Dalitz distributions of three jet events\cite{Brandl}.
\subsection{Study Jet Clustering Algorithms}
Because there are up to four partons in the final state, as many as
three partons can end up in a single jet.  This makes the three jet
calculation sensitive to the details of jet clustering algorithms.
This sort of study in pure gluon production\cite{KG} uncovered an
infrared sensitivity in the commonly used iterative cone algorithms.
\subsection{Study Jet Structure and Shape}
Because there can be three partons clustered into a single jet, this
calculation will allow truly next-to-leading order studies of the
energy distribution in jets.\cite{Shape}  Studies of jet production in
deep inelastic scattering\cite{KRRZ} show that the next-to-leading
order correction for this variable is substantial and agrees rather
well with experimental measurements.
\subsection{Study Event Shape Variables}
There has been a long history of studying event shape variables like
Thrust at $e^+e^-$ colliders.  These measurements challenge the\break
ability of perturbative QCD to describe the data and provide a
means (other than event rate) of obtaining a precise measurement of
$\alpha_s$.  It will be interesting to see if one can make a
meaningful study of such variables at hadron colliders.
\subsection{Study Backgrounds to New Physics}
Models of physics beyond the standard model typically involve massive
states that can generate three jet signals either by associated
production or by decay into three jets.\  To\break identify such signals, one
must understand the pure QCD contribution to three jet production and
look for deviations from the expected distributions.

\section{Results}
At present we have results for some of the most basic event
distributions: the transverse energy spectrum and the angular
distributions.
\begin{figure*}
\epsfxsize30pc
\figurebox{}{}{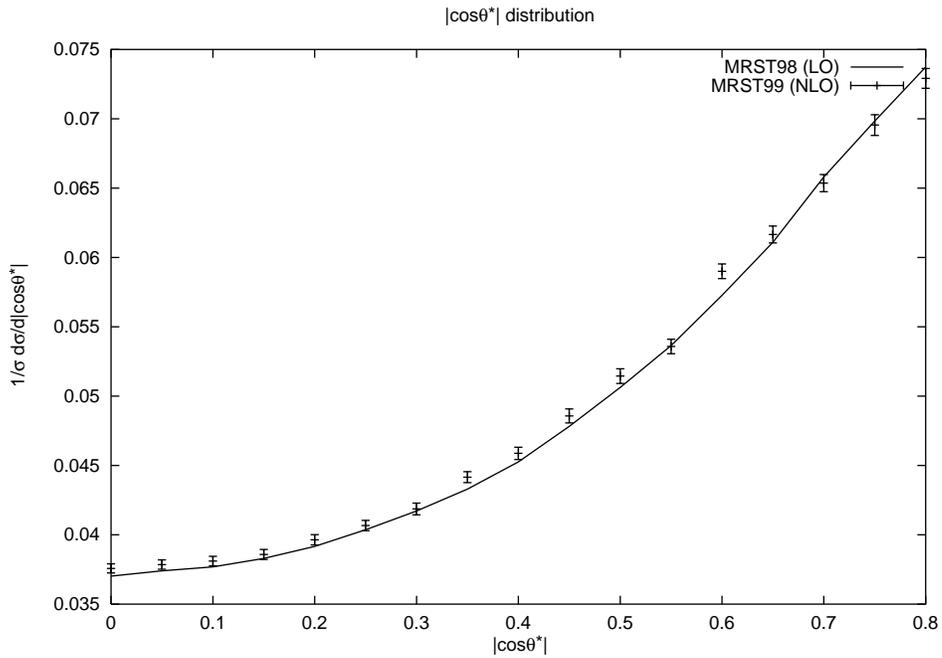}
\caption{
Distribution of events in $\cos\theta^*$, where $\theta^*$ is the
angle between the leading jet and the beam axis in the three jet
center of momentum.  The next-to-leading order results are shown as
points and the leading order results as solid lines.
\label{fig:thetastarplot}}
\end{figure*}
\begin{figure*}
\epsfxsize30pc
\figurebox{}{}{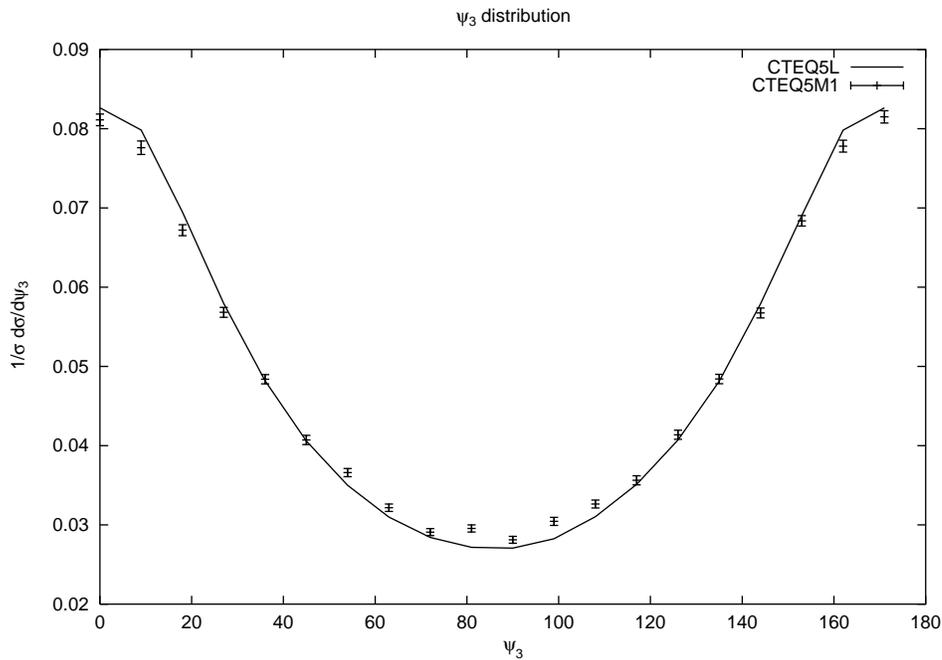}
\caption{
Distribution of events in $\psi_3$, where $\psi_3$ is the
angle between the plane formed by leading jet and the beam axis and
the plane formed by the second and third leading jets in the three jet
center of momentum.  The next-to-leading order results are shown as
points and the leading order results as solid lines.
\label{fig:psithreeplot}}
\end{figure*}
The transverse energy distribution and its scale
dependence has been shown in other conference proceedings\cite{KG2}.
We find that the next-to-leading order correction to the total rate is
very small.  The scale dependence of the NLO calculation, however, is
a factor of two smaller than that of the LO calculation.  The
combination of a small NLO correction and reduced scale dependence
indicates that we are obtaining a reliable calculation of three jet
production.

We also have results for the angular distributions of the three jet
events.  Shown below are the distributions in $\cos\theta^*$, where
$\theta^*$ is the angle between the leading jet and the beam axis in
the three jet center of momentum, and $\psi_3$, where $\psi_3$ is the
angle between the plane formed by leading jet and the beam axis and
the plane formed by the second and third leading jets in the three jet
center of momentum\cite{Multijets}.

Again, the NLO correction is quite small.  The important feature,
however, is that the NLO results are more reliable than the LO
results.  Such distributions are particularly important for
identifying (or eliminating) signals of new physics.  It was the
fact that the angular distributions of dijet events looked like QCD
that eliminated the more exotic explanations of the famous high $E_T$
anomaly in the one jet inclusive distribution.\cite{CDF,D0}

\section*{Acknowledgments}
This work was supported by the US Department of Energy under grant
DE-AC02-98CH10886.

\def\Journal#1#2#3#4{{#1} {\bf #2}, #3 (#4)}

\def\NCA{\em Nuovo Cimento}
\def\NIM{\em Nucl. Instrum. Methods}
\def\NIMA{{\em Nucl. Instrum. Methods} A}
\def\NPB{{\em Nucl. Phys.} B}
\def\PLB{{\em Phys. Lett.}  B}
\def\PRL{\em Phys. Rev. Lett.}
\def\PRD{{\em Phys. Rev.} D}
\def\ZPC{{\em Z. Phys.} C}

\end{document}